\documentclass[useAMS,usenatbib,a4]{mn2e}
\usepackage[dv ips]{graphicx}

\title[The effect of massive binaries on stellar populations and supernova progenitors]{The effect of massive binaries on stellar populations and supernova progenitors}
\author[John J. Eldridge, Robert G. Izzard \& Christopher A. Tout]{John J. Eldridge$^{1,3}$ \thanks{E-mail: jje@ast.cam.ac.uk}, Robert G. Izzard$^{2}$ \& Christopher A. Tout$^{3}$  \\
$^{1}$Astrophysics Research Centre, Physics Building, Queen's University, Belfast, County Antrim, BT7 1NN\\
$^{2}$Sterrekundig Instituut Utrecht, Postbus 80000, 3508 TA Utrecht, The Netherlands\\
$^{3}$Institute of Astronomy, The Observatories, University of Cambridge, Madingley Road, Cambridge, CB3 0HA\\}

\pagerange{\pageref{firstpage}--\pageref{lastpage}} \pubyear{2005}
\begin{document}
\maketitle
\label{firstpage}

\begin{abstract}
We compare our latest single and binary stellar model results from the Cambridge STARS code to several sets of observations. We examine four stellar population ratios, the number of blue to red supergiants, the number of  Wolf-Rayet stars to O supergiants, the number of red supergiants to Wolf-Rayet stars and the relative number of Wolf-Rayet subtypes, WC to WN stars. These four ratios provide a quantitative measure of nuclear burning lifetimes and the importance of mass loss during various stages of the stars' lifetimes. In addition we compare our models to the relative rate of type Ib/c to type II supernovae to measure the amount of mass lost over the entire lives of all stars. We find reasonable agreement between the observationally inferred values and our predicted values by mixing single and binary star populations. However there is evidence that extra mass loss is required to improve the agreement further, to reduce the number of red supergiants and increase the number of Wolf-Rayet stars.
\end{abstract}


\section{Introduction.}
There are many problems outstanding in the field of stellar evolution. The most massive stars, those that end their lives as a supernova (SN), the main questions are `At what rate do stars lose mass?', `Which stars give rise to which SNe?', `What physical processes occur in binary systems?' and `How important are rotation and magnetic fields?' One way to gain insight into these questions is to observe populations of stars and SNe. Massive stars are classified in a range of possible spectroscopic stellar types. The main categories are blue supergiants (BSGs), red supergiants (RSGs) and Wolf-Rayet (WR) stars. If we count the numbers of such stars in a single system (e.g. a stellar cluster or galaxy) we can test the accuracy of our stellar models by comparing to model predictions.

When such comparisons are performed the agreement can be poor. For example the ratio of the BSG to RSG populations has been an outstanding problem for some time \citep{red2blue,mm2001,mo2003}. The ratio of the number of RSGs to the number of WR stars is also not correctly predicted by stellar models \citep{massey}. However the ratios of WR star subtypes WN and WC stars are predicted by modern stellar evolution \citep{mm2005,evink}.

Massive single stars end their lives as core-collapse SNe but the relation between the star and SN type is not straight forward. A list of core-collapse SN types includes IIP, IIL, IIb, IIn, IIpec, Ib and Ic\footnote{Type Ia SNe are thought to be thermonuclear events from explosive carbon burning in a degenerate carbon-oxygen white dwarf and are not considered in this work.} \citep{snclassy}. The difference between type II and type I SNe is the presence or absence of hydrogen in the SN spectrum. It is easy to identify which stellar models contain hydrogen or not but the subtype classifications are fairly arbitrary \citep{popov93,DT03,H03,ETsne,iilafterglow}. The most direct method to link supernovae to stellar end point models is to search for the progenitor stars of SNe in pre-explosion images \citep{VD03,SJM03}.

If we only split the SN types into II and Ib/c another comparison can be made between models and observations. The relative rates of these two SN types and how they vary with metallicity have been estimated observationally \citep{snevsZ}. \citet{mm2005} have shown that standard non-rotating single star models do not agree with these observed rates and that mass loss must be enhanced, in their case by rotation, to find an agreement between predicted SN ratios and those inferred from observations.

The studies mentioned so far only compare single star models to the observations. This ignores the observed fact that many stars are in binaries. Binary evolution has been studied widely, (e.g. \citealt{podsibin1,langerbin,belcz,izzysne,vanbevwr,vaz07,DT07}). The methods employed to estimate the effect of binaries is based either on rapid population synthesis or a small number of detailed models. Population synthesis employs formulae or tables based on detailed models to follow the evolution, (e.g. \citealt{HPT00} and the evolution of a star can be calculated in a fraction of a second rather than many minutes taken by a detailed code. The speed of calculations means that the full range of parameters that govern the outcome of binary evolution can be studied and their relative effects evaluated. The disadvantage is that in the complex phases of evolution, such as when the envelope is close to being stripped, the results of population synthesis can be quite spurious.

We have constructed a set of binary stellar models covering a wide range of binary parameter space using a detailed stellar evolution code \citep{E03}. With these models we have investigated the effect binary evolution has on the lifetimes of the various phases of stellar evolution and therefore the effect it has on the relative populations of massive stars and the relative numbers of different types of SNe. While the binary parameter-space resolution is low compared to rapid population synthesis studies it is still one of the highest resolution studies undertaken with a detailed evolution code.

We note that we do not consider here the effect of rotation or magnetic fields. The effect of rotation on the structure and evolution of stars is a complicated process and has been studied by \citet{hegrot1} and \citet{hegrot2} and the extensive study beginning in \citet{mm1997} and continuing up to most recently in \citet{mmgrb}. The effect of magnetic fields on stellar structure and evolution has also been investigated by \citet{mm2003}, \citet{mm2004}, \citet{hegermag}. Our code does not include rotation therefore our study reveals the importance of duplicity alone. We discuss how rotation may affect our results in Section 6.

In this paper we start by providing details of our stellar model construction. We then specify our definitions of the different phases of stellar evolution and discuss how binary evolution affects the relative lifetimes of stars in various phases of stellar evolution. Using these lifetimes we predict how the BSG to RSG, the WR to O supergiant, the RSG to WR and WC to WN ratios vary with metallicity. We then predict how the relative rates of type Ib/c and type II SNe vary with metallicity. With each of these predictions we compare the values to the ratios inferred from observations. Finally we discuss the implications of our results.

\section{Construction and Testing of the Stellar Models.}

We use a detailed stellar evolution code to construct sets of single and binary star models rather than a rapid population synthesis code, because we require accurate treatment of complex stages of evolution, such as when the envelope is close to being stripped, to obtain more accurate pre-SN models. We model the binary parameters with a reasonable resolution owing to the great amount of computer power that is now available. We have compared our binary models to the predictions of a rapid population synthesis code \citep{izzy,izzysne,izzynew} and found reasonable agreement for most of our predictions. However difficulty in assigning WR types to population synthesis helium stars meant the agreement was poor for the WC/WN ratio.

\subsection{Detailed Single Star Models}

Our detailed stellar models were calculated with the Cambridge stellar evolution code, STARS, originally developed by \citet{E71} and updated most recently by \citet{P95} and \citet{E03}. Further details can be found at the code's web pages\footnote{\texttt{http://www.ast.cam.ac.uk/$\sim$stars}}. Our models are available from the same location for download without restriction. They are similar to those described by \citet{ETsne} but we use 46 rather than 21 zero-age main-sequence models. Every integer mass from $5$ to $40{\rm M}_{\odot}$ is modelled along with 45, 50, 55, 60, 70, 80, 100 and $120{\rm M}_{\odot}$. The initially uniform composition is $X=0.75-2.5Z$ and $Y=0.25+1.5Z$, where $X$ is the mass fraction of hydrogen, $Y$ that of helium and $Z$ the initial metallicity taking the values 0.001, 0.004, 0.008, 0.02 and 0.04. The composition mixture is scaled solar and ${\rm Z}_{\odot}=0.02$. Our models end after core carbon-burning or at the onset of core neon ignition. During these burning stages, the envelope is only affected to a small degree because $t_{\rm nuclear} \ll t_{\rm thermal}$. Furthermore the late stages of evolution are rapid and have a negligible effect on the final observable state of the progenitor \citep{WHW02}.

Our opacity tables include the latest low temperature molecular opacities of \citet{FA05}. This update leads to tiny changes in the radii and surface temperatures of red supergiant models because the altered opacity values only slightly modify the surface boundary conditions.

All the models employ our standard mass-loss prescription because it agrees best with observations of SN progenitors \citep{ETsne}. The prescription is based upon that of \citet{DT03} but modified in several ways. For pre-WR mass loss we use the rates of \citet{dJ} except for OB stars for which we use the theoretical rates of \citet{VKL2001}. When the star becomes a WR star ($X_{\rm surface}<0.4$, $\log (T_{\rm eff}/{\rm K})>4.0$) we use the rates of \citet{NL00}. We scale all rates with the standard factor $(Z/$Z$_{\odot})^{0.5}$ \citep{K87,H03}, except for the rates of \citet{VKL2001} which include their own metallicity scaling.

The most important change is that we scale the WR mass-loss rates with the by metallicity, even though this is not normally included. \citet{vanbmassloss} first included the scaling in a population synthesis model while \citet{WRZscale} first suggested that the scaling should be included from observations of WR stars in the LMC and in the Galaxy. Recent theoretical predictions also suggest the scaling should be included \citep{Vink2005}. Inclusion of this scaling results in a greater agreement with the change of WR population ratios with metallicity \citep{evink}. There is some uncertainty in the exact magnitude of the exponent we should use. To remain consistent we use the same exponent as for the non-WR stars.

We synthesise a single star population by assuming a constant star formation rate and the initial mass function (IMF) of \citet{KTG93}.

\subsection{Detailed Binary Models}

Binary stars experience different evolution from that of single stars if their components interact. In a wide enough orbit a star is not affected by a companion. Duplicity allows the possibility of enhanced mass loss, mass transfer and other binary specific interactions (e.g. irradiation, colliding winds, surface contact, gravitational distortion) and hence a greater scope for interesting evolutionary scenarios. 

We have modified our stellar evolution code to model binary evolution. The details of our binary interaction algorithm are relatively simple compared to the scheme outlined in \citet{HPT02}. We use their scheme as a basis but we change some details which cannot be directly applied to our detailed stellar evolution calculation. We also make a number of assumptions in producing our code to keep it relatively simple. Our aim is to investigate the effect of enhanced mass-loss due to binary interactions on stellar lifetimes and populations, therefore we concentrate on this rather than every possible physical process which would add more uncertainty to our model. We also make assumptions in calculating our synthetic population to avoid calculating a large number of models. For example we do not model the accretion on to the secondary in the detailed code. We take the final mass of the secondary at the end of the primary code as the initial mass of the secondary when we create our detailed secondary model. This avoids calculating ten times more secondary models than primary models.

\subsubsection{Evolution of a binary}

In our binary models we always treat the primary as the initially more massive star when and we only evolve one star at a time with our detailed code. When we evolve the primary in detail it has a shorter evolutionary timescale than the secondary which remains on the main sequence until after the primary completes its evolution and so we can determine the state of the secondary using the single stellar evolution equations of \citet{HPT00}. When we evolve the secondary in detail we assume its companion is the compact remnant of the primary (a white dwarf, neutron star or black hole) and treat this as a point mass. We describe how we combine these models to synthesise a population below.

To model a binary we begin by specifying a primary mass, $M_{1}$, a secondary mass, $M_{2}$, and an initial separation, $a$. We assume the orbit is circular and conserve angular momentum unless mass is lost from the system. We follow the rotation of the stars, considering them to rotate as solid bodies but we do not include processes such as rotationally-enhanced mass loss or mixing. We give the stars initial rotation rates linked to their initial masses as in \citet{HPT00}. When the stars lose mass in their stellar wind angular momentum is lost both from their spin and orbit. We ignore tides so stellar rotation and orbital rotation evolve independently until Roche-lobe overflow (RLOF) when we force the stars into synchronous rotation with the orbit, transferring angular momentum from the spin to the orbit or vice versa.

\subsubsection{Roche-lobe overflow}

In a binary system the equipotential surfaces are not spherical but are distorted by the gravity of the companion star and by the rotation of the system. If the stars are small compared to the size of the orbital separation these surfaces are nearly spherical but as the star grows relative to the separation the shape is distorted and becomes more ellipsoidal. A star's surface eventually reaches the L1 Lagrange point where the gravity of both stars cancels exactly. If a star expands beyond this point then material begins to flow towards the other star. To include this in our models we define a Roche lobe radius, $R_{\rm L1}$, such that the sphere of this radius has the same volume as the material within the Roche Lobe defined by the equipotential surface passing through the L1 point. Therefore when the star has a radius greater than the Roche lobe radius it transfers material on to the other star. We use the Roche lobe radius given by \citet{E83},
\begin{equation}
\frac{R_{\rm L1}}{a}=\frac{0.49 q_{1}^{2/3}}{0.6 q_{1}^{2/3}+\ln (1+q_{1}^{1/3})},
\end{equation}
where $q_{1}=M_{1}/M_{2}$ and $a$ is the orbital separation. It is accurate to within 2\% for the range $0<q_{1}<\infty$. When $R_{1} > R_{\rm L1}$ we calculate the rate at which mass is lost from the primary according to,
\begin{equation}
\dot{M}_{\rm 1R}=F(M_{1})[\ln (R_{1}/R_{\rm L1})]^{3} {\rm M}_{\odot} {\rm yr^{-1}},
\end{equation}
where
\begin{equation}
F(M_{1}) =3 \times 10^{-6} [\min(M_{1},5.0)]^{2},
\end{equation}
chosen by experiment to ensure mass transfer is steady \citep{HPT02}.

Mass lost from the primary is transferred to the secondary but not all is necessarily accreted. Accretion causes the star to expand owing to increased total mass and therefore an increased energy production rate if $\dot{M}_{\rm 2} \ge  M_{2} / \tau_{\rm KH}$ where $\tau_{\rm KH}$ is the thermal, or Kelvin-Helmholtz, timescale. We assume that the star's maximum accretion rate is determined by its current mass and its thermal timescale. We define a maximum accretion rate for a star such that $\dot{M}_{\rm 2} \le M_{2} / \tau_{\rm KH}$. If the accretion rate is greater than this then any additional mass and its orbital angular momentum are lost from the system. In general stars with lower masses have longer thermal timescales than more massive stars. Efficient transfer is only possible if the two stars are of nearly equal mass so the thermal timescales are similar. This is an approximate treatment but provides a similar result to the more complex model of \citet{petrovic} who included rotation and found it led to inefficient mass-transfer. For compact companions we derive the maximum accretion rate from the Eddington limit \citep{edding}.

\subsubsection{Common-envelope evolution}

If RLOF occurs but does not arrest the expansion of the mass losing star, growth continues until the secondary is engulfed in the envelope of the primary. This is common envelope evolution (CEE). During CEE it is thought that the envelope is ejected by some dynamical process and that the principal source of energy for the envelope ejection is the orbital energy \citep{cee1,cee2}. The orbital separation decreases and there is a chance that the two stars may coalesce before the envelope is ejected. Alternatively the two stars are left in a close orbit, commonly one helium star for a massive progenitor and one main-sequence star. If the secondary then evolves to a helium giant a second CEE phase can occur possibly leading to a merger or a very close binary.

CEE is modelled in population synthesis by first calculating the energy required for the ejection of the envelope, the binding energy, $E_{1,{\rm bind}}=GM_{1}M_{1,{\rm env}}/(\lambda R_{1}$, where $M_{1,{\rm env}}$ is the mass of the envelope, $R_{1}$ is the radius of the star and $\lambda$ is a constant to reflect the structure of the envelope. This is then compared to the initial orbital energy, $E_{{\rm orb},i}=-GM_{c,1}M_{2}/(2a_{i})$, where $M_{c,1}$ is the mass of the core and $a_{i}$ is the initial separation at the onset of CEE. If there is not enough orbital energy to eject the envelope and leave a stable binary then the two stars merge to form one star. Otherwise the envelope is removed leaving a stable binary system (e.g. \citealt{HPT02}). The new orbit is calculated by the equation $E_{1,{\rm bind}} = \alpha_{\rm CE} ( E_{{\rm orb},i}-E_{{\rm orb},f}  )$, where $E_{{\rm orb},f}$ is the final orbital energy from which the final separation is calculated. There is some uncertainty in the constant $\alpha_{\rm CE}$ as there are other sources of energy available such as the energy from nuclear reactions and re-ionisation energy of the hydrogen in the envelope so it can be greater than one.

In our models when CEE occurs we derive the mass-loss rate from the above equation for RLOF, however we limit the mass-loss rates to a maximum of $10^{-3} \, $M$_{\odot} {\rm yr^{-1}}$ because more rapid mass loss causes the evolution code to break down. We assume that the secondary accretes no mass because the CEE occurs on the thermal timescale of the primary which is normally shorter than the thermal timescale of the less massive secondary. The accretion rate for a compact remnant is restricted to the Eddington limit \citep{edding}. To calculate the change in the orbital separation we calculate the binding energy of the material lost in the CEE wind ($\delta E_{\rm binding}=-(M_{1}+M_{2}) \delta M_{1}/R_{1}$) and remove it from the orbital energy ($E_{\rm orbit}=-GM_{1}M_{2}/(2a)$). We find,
\begin{equation}
{\rm \delta}a = \frac{a^{2}}{R_{1}} \frac{M_{1}+M_{2}}{M_{2}} \frac{{\delta}M_{1}}{M_{\rm 1}},
\end{equation}
where $\delta a$ is the change in orbital separation and $\delta M_{1}$ is the mass lost in the time period considered, one model timestep. In this formalism ${\delta}M_{1}$ is negative so the orbit shrinks. By determining the mass-loss rate from the RLOF equation we find that the CEE ends naturally with the stellar radius shrinking once most of the envelope has been removed.

\subsubsection{Binary mergers}

In the most compact binaries the stars can merge during CEE to form a single star whose mass is the sum of its parents. We find that some of our binary models enter CEE and the binary separation tends to zero. When this occurs we use a different binary model. The primary is evolved in the detailed code and once its radius is equal to the binary separation all the mass of the secondary is accreted on to the surface of the primary star and the evolution continues as a more massive single star. This is an extremely simple model. Other models such as \citet{podsibin1} show that a more detailed treatment may be required but the efficiency of the merger is always uncertain. Our models provide an estimate of the effects of stellar mergers and are an upper limit as we assume all the secondaries mass is accreted. Furthermore the outcome depends on the evolutionary state of the primary, it is possible that fresh hydrogen could be mixed into the helium burning zone and therefore cause the envelope to be explosively removed \citep{podsicee}.

\subsubsection{The binary model subsets}

We calculate three subsets of binary models which we then combine to simulate a population of stars. The first subset consists of our primary models where we evolve the primary star in the detailed code and evolve the secondary according to the single-star evolution equation of \citet{HPT00}. We choose our grid to be $M_{1}=$5, 6, 7, 8, 9, 10, 11, 12, 13, 15, 20, 25, 30, 40, 60, 80, 100 and 120; $q=\frac{M_{2}}{M_{1}}=0.1$, $0.3$, $0.5$, $0.7$ and $0.9$; $\log (a/$R$_{\odot}) =$1.0, 1.25, 1.5, 1.75, 2, 2.25, 2.5, 2.75, 3, 3.25, 3.25, 3.5, 3.75 and 4;. We do not model lower masses because these stars are never luminous enough to affect our population ratios and do not lead to core-collapse SNe.

 The second subset consists of our secondary models where we evolve the secondary in the detailed code after the end of the primary evolution and replace the primary by its remnant. We chose three masses for the compact companion to model a white dwarf, neutron star or black hole, $M_{1,{\rm post-SN}}=0.6$, $1.4$ or 2M$_{\odot}$ respectively. We choose the grid to use the same separation and mass grid as for our primary model subset. The final subset consists of our merger models and we choose this grid to have the same $M_{1}$, $a$ and $q$ distribution as our primary model. We only calculate models with initial orbital separation smaller than the maximum stellar radius for each initial primary mass. These are only used if the orbital separation reaches zero in a primary model.

\subsubsection{Synthesising the population}

The binary population is synthesised by first assuming the primary masses are distributed with the IMF of \citet{KTG93}. Then we assume flat distribution over $1 \le \log\, (a/$R$_{\odot}) \le 4$ and $0.1 \le q \le 0.9$. We calculate the primary stars' lifetimes and SN types using the primary models. If a primary model during evolution enters a CEE and the binary separation shrinks to zero we replace the primary model with a merger model.

The end point of the primary model is taken to be at the onset of neon burning, or the onset of thermal pulses in an AGB star. First we determine the remnant the primary will form. A white dwarf is selected unless the primary underwent a SN if there is an oxygen-neon core and the carbon-oxygen core mass is greater than 1.3 M$_{\odot}$ when a neutron star is selected. A black hole remnant replaces this if the helium core mass is greater than 8 M$_{\odot}$. In the case of a white dwarf the system remains bound. When a neutron star is formed we consider the effect of its natal kick on the binary orbit. We use the prescription of \citet{nskick} and the kick distribution of \citet{kick2} to determine whether the binary remains bound or not. In the black hole case we assume the kick velocity is zero and estimate the remnant mass and ejected mass as in \citet{ETsne}.

If the system is unbound we calculate the contribution of the secondary to the  stellar lifetimes and SN type from our single star models. The contribution from binaries that remain bound are taken from our secondary models. The initial mass taken for the secondary star is the initial secondary mass unless the star accreted material from the primary in which case we use the new mass after mass-transfer.

Combining the model subsets creates our synthetic population of binary stars based on detailed models from which we can estimate the relative numbers of different stellar types and the relative rates of different types of SNe. We calculate these populations at five different metallicities, $Z=0.001$, 0.004, 0.008, 0.02 and 0.04. 

\subsubsection{Uncertainties}

Uncertainties arise from assumptions we have made in the physical model for binary systems. The greatest uncertainty is that we do not model the primary and secondary in detail simultaneously. This means we are not able to model systems where $q$ is close to one. In these mass transfer can be quite efficient with the secondary accreting a large fraction of the mass of the primary \citep{bingrb}. Our approximation is reasonable because for primaries with initial masses $\la 60M_{\odot}$, the secondary stars with $q \la 0.9$ remain on the main-sequence during the primary lifetime.  The primaries $>60M_{\odot}$ with companions that evolve off the main sequence before the primary experiences a SN also rarely interact as both stars lose the hydrogen envelopes before becoming RSGs. The error introduced by this approximation is dwarfed by the uncertainty in our assumed initial parameter distribution for the binaries. If we only consider binaries closer than $\log (a/R_{\odot}) < 3$ then we find that our RSG/WR ratio decreases by at least a factor of two.

%%%%%%%%%need to change to also say those that do evolve of main sequence tend to be most massive stars that never become RSGs anyway

\subsection{Defining Stellar Types}

\begin{table}
\caption{Our definitions of stellar types.}
\label{stellartypes}
\begin{tabular}{@{}lcccc@{}}
\hline
Name			& $\log(T_{\rm eff}/{\rm K})$	& $\log(L/{\rm L_{\odot}})$	&	$X$	&	$\frac{(x_{\rm C}+x_{\rm O})}{y}$ \\
\hline
Blue supergiants	&	$\ge 3.9$		&	$\ge 4.9$		&$> 0.4$	&	-	\\	
Red supergiants 	&	$\le 3.66$		&	$\ge 4.9$		&$> 0.001$	&	-	\\
O supergiants           &       $\ge 4.48$              &       -                       &$> 0.4$        &       -       \\
\hline
Wolf-Rayet star		&	$\ge 4.0$		&	$\ge 4.9$		&$\le 0.4$	&	-	\\
WNL star		&	$\ge 4.0$		&	$\ge 4.9$		&$\le 0.4$	&	-  \\
WNE star		&       $\ge 4.0$		&	$\ge 4.9$		&$\le 0.001$	&	$\le 0.03$ \\
WC star			&	$\ge 4.0$		&	$\ge 4.9$		&$\le 0.001$	&	$> 0.03$ \\
\hline
\end{tabular}
\end{table} 

\begin{figure}
\includegraphics[angle=0, width=84mm]{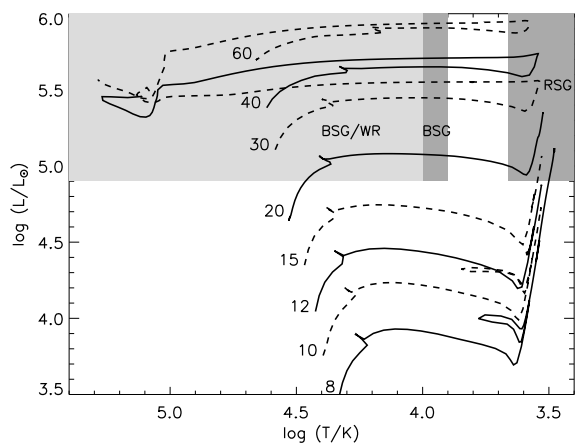}
\caption{Hertzsprung-Russel diagram indicating regions that we include as BSGs, RSGs and WR stars with single star tracks to indicate the mass ranges they sample. The numbers indicate the initial stellar mass in M$_{\odot}$ for each track. The initial metallicity was Z=0.02. The BSG and WR regions overlap in the lighter shaded region adjacent to the BSG region. The distinction between WR and BSGs is made by surface hydrogen abundance.}
\label{hrexample}
\end{figure}

Stellar types may be defined by observable characteristics such as colour and luminosity. However for a complete and accurate classification spectra are required. For example some objects may initially appear to be BSGs but their spectra indicate they are quite different and are in fact WR stars. WR spectra contain broad emission lines and weak or no hydrogen lines indicative of fast and dense stellar winds. In this work we use the theoretical stellar definitions that are summarised in Table \ref{stellartypes}.

First, we only consider massive BSGs, RSGs and WR stars that have $\log (L/{\rm L}_{\odot}) \ge 4.9$, equivalent to a luminosity cut-off of $M_{\rm bol} \ge -7.5$. If we compare our models to single clusters with well defined ages the limit would not be critical. However, because we are considering a less specific population with an assumed constant star-formation rate, lowering the limit slightly to $\log (L/{\rm L}_{\odot}) \ge 4.7$ does not significantly change our predicted BSG to RSG and RSG to WR ratios but does decrease the WC to WN ratio because a larger number of low-mass helium stars are included as WR stars.

BSGs and RSGs both have hydrogen envelopes but differ in their surface temperatures. BSGs are hot with $\log(T_{\rm eff}/{\rm K}) \ge 3.9$. This includes O, B and A stars. RSGs are cooler with $\log(T_{\rm eff}/{\rm K}) \le 3.66$. This includes K and M stars.  We show how these regions fit the tracks of our single star evolution models in Figure \ref{hrexample}. RSGs are stars more massive than 12M$_{\odot}$ while BSGs are stars more massive than 20$M_{\odot}$. Our luminosity limit reduces contamination from AGB stars \citep{mo2003} in the RSG sample. The limit is also greater than the luminosity of stars that experience a blue loop during helium burning. This feature is very sensitive to the details of convection included in the stellar models and so best excluded \citep{red2blue}. Our models with blue loops lie outside of our RSG definition so our predicted ratios are less sensitive to the complications introduced by the details of convection and mixing employed in our models.

We also consider the O supergiants (OSGs) which are a subset of the BSGsfor the WR/O ratio. OSGs are the hottest hydrogen-rich stars with $\log(T_{\rm eff}/{\rm K}) \ge 4.48$ and a surface hydrogen mass fraction, $X>0.4$. There is no luminosity limit for these stars as in \citet{mm2005}.

The other stellar types we consider are WR stars. We define a WR star to be any star that has $\log(T_{\rm eff}/{\rm K}) \ge 4.0$ and a surface hydrogen mass fraction, $X < 0.4$. We also require that $\log (L/{\rm L}_{\odot}) \ge 4.9$. Further to this we subdivide WR stars into WNL, WNE and WC stars. In each case the element nitrogen or carbon is dominant in their emission spectrum. The sequence is due to the exposure of nuclear burning products on the surface of the WR stars. We assume a WR star is initially a WNL star. When $X < 0.001$ it becomes a WNE star. The star becomes a WC star when $X < 0.001$ and $ (x_{\rm C}+x_{\rm O})/y > 0.03$ where $x_{\rm C}$, $x_{\rm O}$ and $y$ are the surface number fractions of carbon, oxygen and helium.

\section{Blue and Red supergiant and Wolf-Rayet lifetimes.}

In this section we compare the stellar lifetimes of our single and primary models at two different metallicities. We calculate a mean lifetime from our primary models only to represent the effect on the lifetimes of RLOF and CEE. The mean lifetime for primary models with the same initial mass is calculated assuming a flat distribution in $q$ and $\log a$.

The primary model BSG mean lifetimes are slightly increased relative to the single star BSG lifetime. The total average lifetime for BSGs increases by 4 to 8 per cent and there is no trend with metallicity. This is because close merger systems extend the burning lifetimes of the primary star. Also stars initially too low in mass to be counted as BSGs accrete enough mass to rise above our luminosity limit.

\begin{figure}
\includegraphics[angle=0, width=84mm]{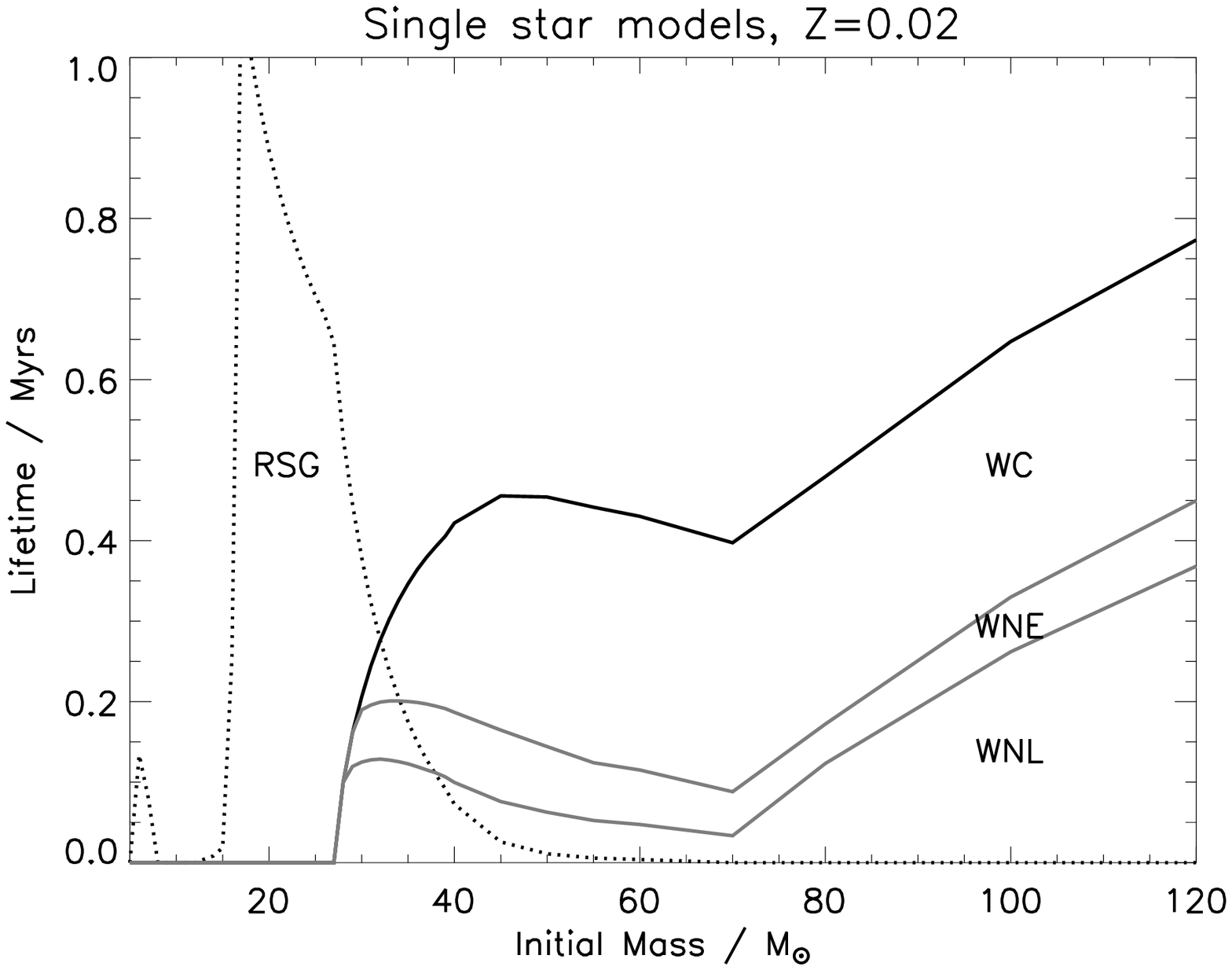}
\includegraphics[angle=0, width=84mm]{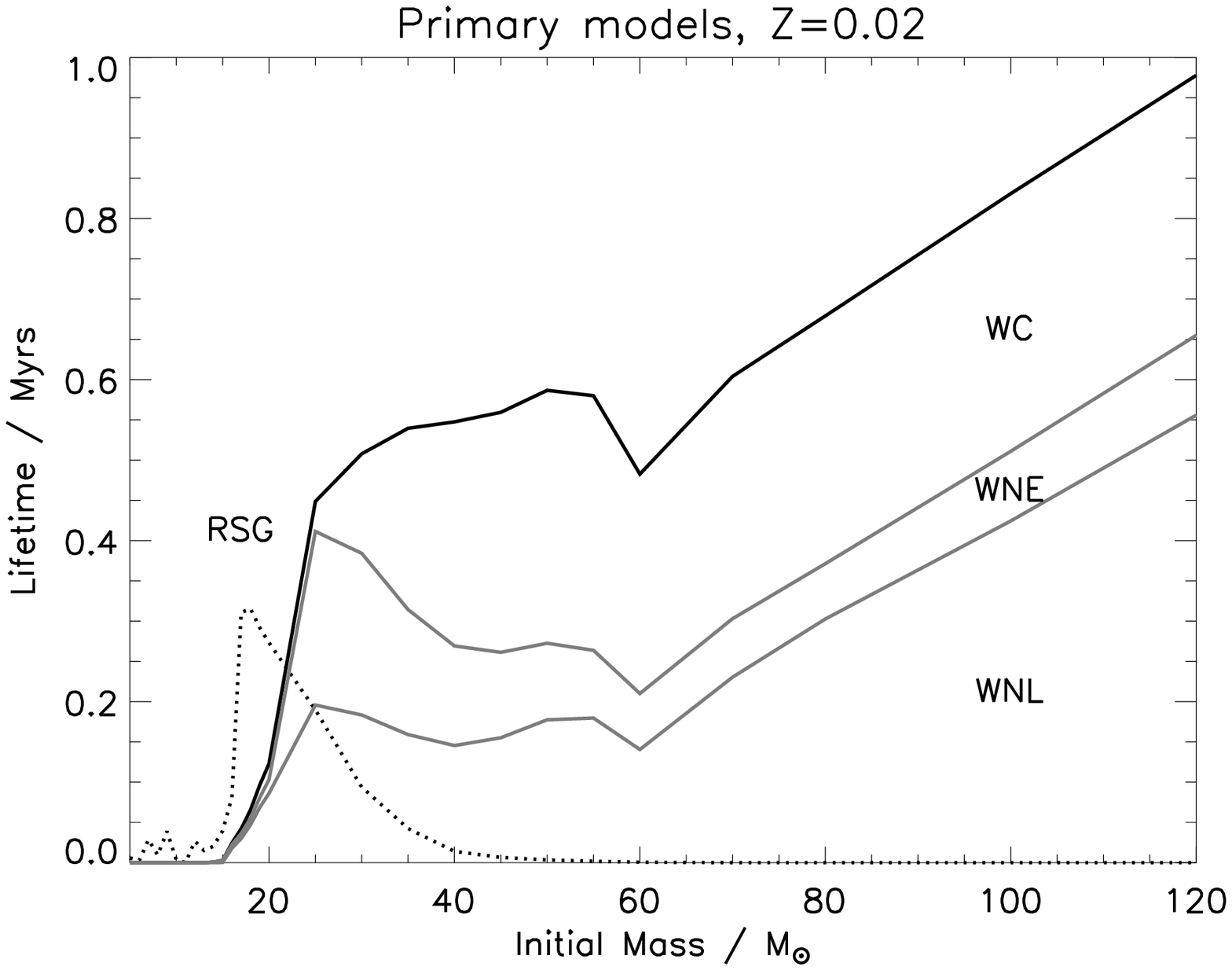}
\caption{RSG, WNL, WNE and WC lifetimes at Z=0.02. Upper panel, our single star models. Lower panel, our primary models. The lifetimes of the binary stars are mean lifetimes assuming a flat $q$ and $\log a$ distribution. Each region is labelled but progresses from bottom to top by WNL, WNE and WC phase.}
\label{RSGS}
\end{figure}

\begin{figure}
\includegraphics[angle=0, width=84mm]{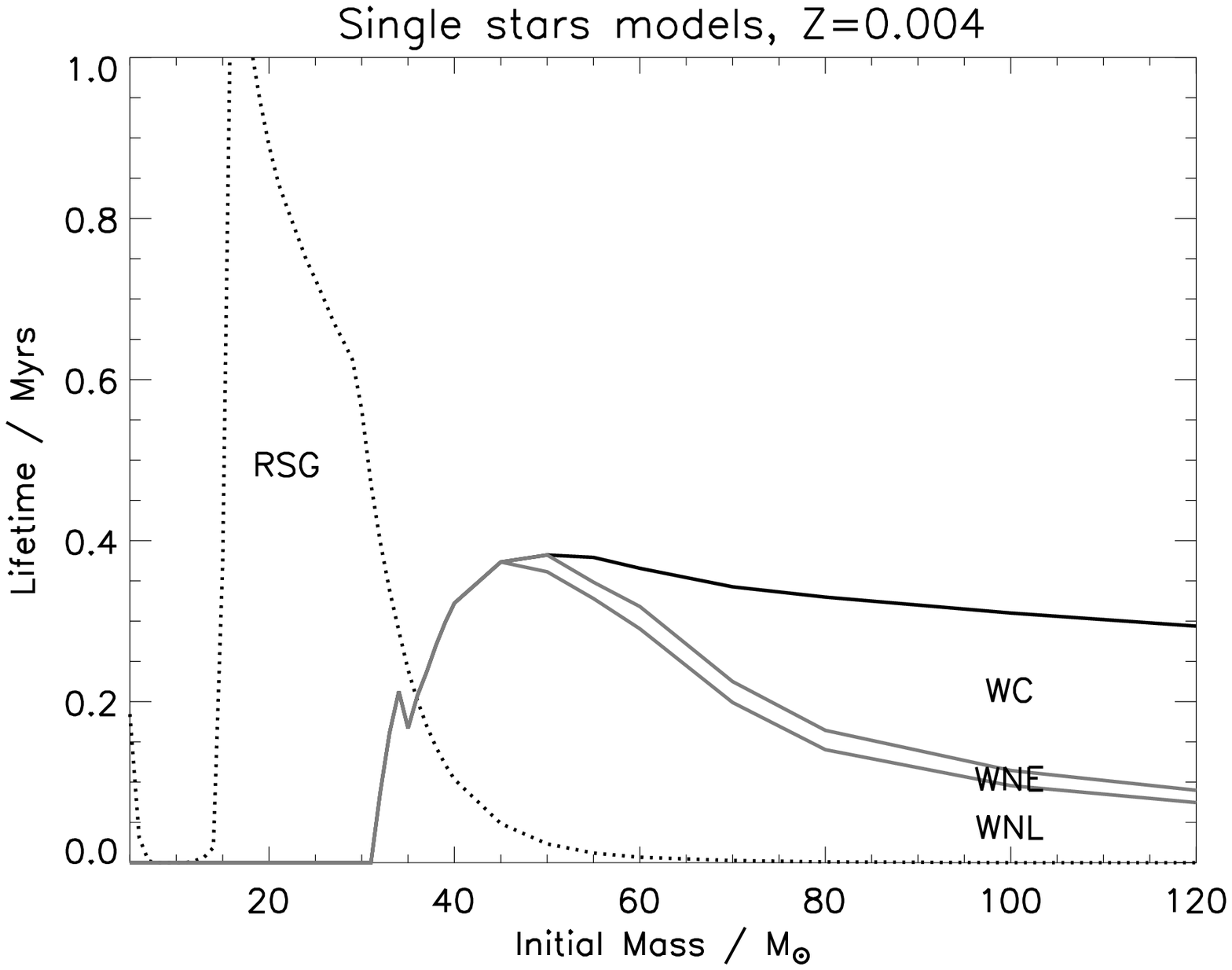}
\includegraphics[angle=0, width=84mm]{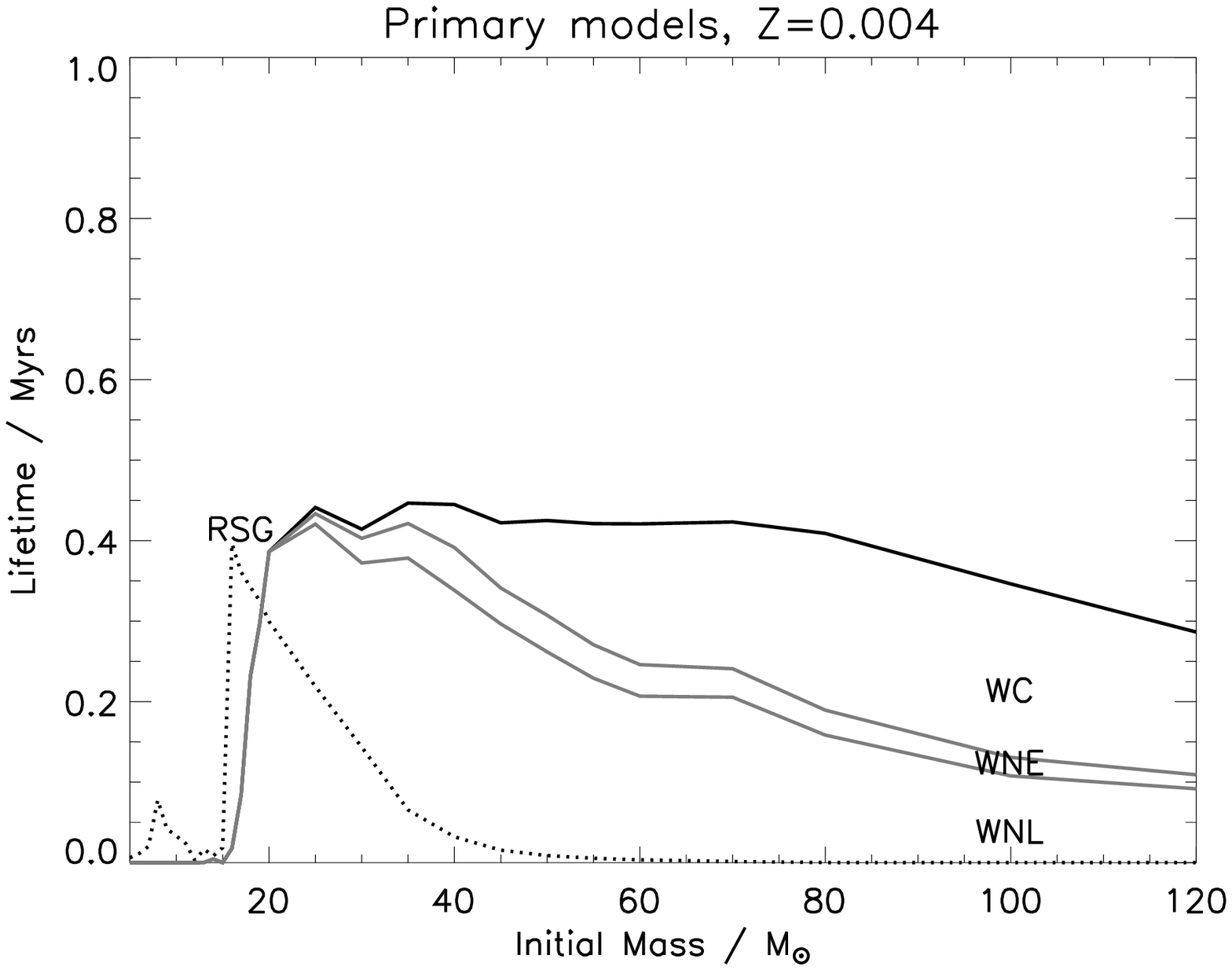}
\caption{As for Figure \ref{RSGS} but Z=0.004.}
\label{WNLS}
\end{figure}

The RSG lifetimes of our models are shown by the dotted lines in Figures \ref{RSGS} and \ref{WNLS}. The single star lifetimes at $Z=0.004$ are longer than those at $Z=0.02$. This is due to the reduction in mass-loss rates at the lower metallicity so more time is required for the hydrogen envelope to be removed. Our primary models have RSG lifetimes two to three times shorter. RLOF and CEE greatly reduce the number of RSGs by removing their hydrogen envelopes to form WR stars.

The WR lifetimes of our single stars are identical to those of \citet{evink}. Mass loss affects the lifetimes. Firstly, it determines the {\it total} WR lifespan by affecting the total stellar mass and thus the vigour of the nuclear reactions in the core. Secondly, stronger mass loss strips mass more quickly from the stellar surface, which leads to the exposure of hydrogen and then helium burning products and results in the appearance of different WR subtypes. 

The WR lifetimes are longer at higher metallicities as the WR phase is entered at an earlier stage and there is a longer period of core helium burning during the WR phase because of larger pre-WR mass loss. The kink seen in the Z=0.02 plot (Fig. \ref{RSGS}) is due the fact that stars with masses below it undergo a red supergiant phase when mass is stripped and the surface hydrogen abundance drops below the limit set for WNL stars ($X_{\rm surface}$ $\le$ $0.4$). However, the stars are still cool and some time passes before they are hotter than the temperature limit ($10^{4}\, {\rm K}$). Stars more massive than the mass at the kink spend most of their evolution at temperatures hotter than this limit and become WR stars as soon as the surface abundance requirement is reached.

Our primary model WR lifetimes in Figures \ref{RSGS} and \ref{WNLS} have relatively minor differences as far as the total WR lifetimes are concerned. The main difference is that the minimum mass for a WR star is decrease to around $15M_{\odot}$ at both metallicities. These stars will produce lower mass WR stars than are created by single stars.

The lifetimes of the WR subtypes at $Z=0.02$ indicate that the primary models have longer WNE lifetimes than single stars and shorter WC lifetimes. The WR stars below the minimum mass limit for single WR stars do not have a WC phase and exist only as WN stars. At $Z=0.004$ we see the same trend but many of the lower mass WR stars only have WNL phases.

\section{Ratios of Stellar Number Counts}

Counting and comparing the number of stars in different stellar populations is a useful method for investigating stellar evolution. A comparison between observed populations and those predicted by a stellar evolution code is only worthwhile if in the count stars are not missed or misclassified. Many older observations relied on groups of stars determined by photometry. The use of spectrometry has improved the accuracy and completeness of observational surveys \citep{mo2003}.

The metallicity mass fraction of the observed stellar populations was calculated from the $\log[{\rm O/H}]+12$ values and comparing them to the values from our sets of models. Because this process is ambiguous and the position of solar metallicity for the mass-loss scaling is indistinct, we have assumed the metallicities are uncertain by 25 per cent. 

\subsection{Blue to Red Supergiant Ratio}

The ratio of the number of BSGs to the number of RSGs has been a problem in stellar astrophysics for some time \citep{red2blue}. Observations tend to indicate that the BSG/RSG ratio should decrease with metallicity while most stellar evolution codes predict a constant or increasing ratio. \citet{EMM} suggested that the situation could be improved by using spectroscopy to confirm the identity of supergiants. If we compare our models to their observed trend we find good agreement at solar metallicity but no agreement at SMC metallicity. However \citet{chiosi} indicate that the SMC cluster \citet{EMM} observed has a spread of ages similar to the age of the cluster itself, therefore the ratio they observe is not due to a single population of stars with the same age.

\begin{table}
\caption{Observed numbers and population ratio of BSGs and RSGs. Values are taken from \citet{mo2003}.}
\label{blue2redtable}
\begin{tabular}{@{}lcccc@{}}
\hline
System	&$\log [{\rm O/H}]+12$	&$N_{\rm BSG}$	&$N_{\rm RSG}$	&$N_{\rm BSG}/N_{\rm RSG}$	\\	
\hline
SMC	&8.13		&1484	&90	&$16 \pm 6$		\\
LMC	&8.37		&3164	&234	&$14 \pm 4$		\\
\hline
\end{tabular}
\end{table} 

\begin{figure}
\includegraphics[angle=270, width=84mm]{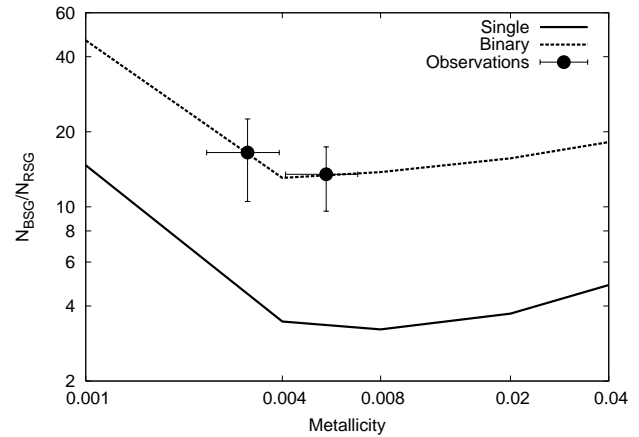}
\caption{Ratio of the numbers of BSGs to RSGs versus metallicity. Observations are taken from \citet{mo2003}. The solid line is from our single star models while the dashed line is from our binary models.}
\label{blue2red}
\end{figure}

\citet{mo2003} provide a detailed observational study of RSGs in the Magellanic clouds with rigid definitions for BSGs and RSGs (see Table \ref{stellartypes}). They found from observing the entire SMC and LMC that the ratios of BSGs to RSGs were 16 and 14 respectively (see Table \ref{blue2redtable}).

Comparing the observations to predicted values from our models in Figure \ref{blue2red} we find that single stars predict too low a value because either there are too many RSGs, too few BSGs or a combination of these two. Extra BSGs can be made via stellar mergers. Such stars are seen in globular clusters when the mass of the merged star is greater than the mass of the cluster turn-off, these are blue stragglers \citep{bluestrag}. The second binary effect is that stars below the minimum initial single star mass for a WR star will lose their hydrogen envelopes to become WR stars because of RLOF or CEE. These are stars that would have been RSGs and therefore the total number is reduced. In combination these processes increase the predicted BSG/RSG ratio as our binary models show.

The binary model ratio is dependent on how we choose the distribution of binaries in $q$ and $\log \, (a/{\rm R}_{\odot})$. In our binary population we include very wide binaries, $\log \, (a/{\rm R}_{\odot}) \ge 3$, which evolve as single stars so our binary population is a mix of interacting binaries and single stars. Approximately one third of our primary models evolve as single stars. Therefore, to reproduce observations, two thirds of all stars must be in interacting binaries. If our binary models are not completely correct something extra, such as rotation, may decrease the required fraction of interacting binaries. \citet{mm2001} find that rotation has a strong effect on the BSG to RSG ratio but the trend they found was that the BSG to RSG ratio decreased rather than increased.% However, tides may change this by affecting the rotation of the stars.

\subsection{Wolf-Rayet to O supergiant Ratio}

The WR to OSG ratio has been studied for sometime \citep{mm1994}. If binaries leave the RSG population to become WR stars therefore this ratio should increase if the BSG to RSG ratio decreases. The observed ratio is less certain than the BSG to RSG ratio because there is greater uncertainty in the completeness of the observations \citep{massey}. However the trend that the ratio decreases with metallicity is well established \citep{crowther2007}.

\begin{figure}
\includegraphics[angle=270, width=84mm]{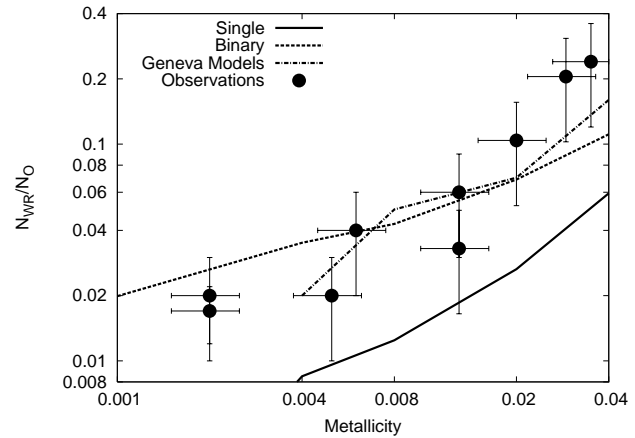}
\caption{Ratio of the numbers of Wolf-Rayet stars to O-supergiants versus metallicity. Observations are taken from \citet{mm1994}. The solid line is our single star models while the dashed line is our binary models. The dashed-dotted line for the Geneva models is taken from \citet{mm2005}. The y-axis error bars are an assumed error of 50 percent of the values given by \citet{mm1994}.}
\label{blue2wr}
\end{figure}

Figure \ref{blue2wr} shows that our single star models underpredict the ratio while our binary models are in better agreement, as are the Geneva rotating models. This is because in our models more stars are stripped of their envelopes and become WR stars. There is also a contribution to the OSG population from stellar mergers and secondary accretion as for the BSG/RSG ratio above. The trend our binary models predict with metallicity is a little too shallow but within the uncertainty of the observed ratios. This again could indicate something extra may need to be included in our models.

\subsection{Red Supergiant to Wolf-Rayet Ratio}

The RSG to WR ratio compares the relative populations of the two stellar types that have completed core hydrogen burning and so measures the influence of mass loss.

\begin{figure}
\includegraphics[angle=270, width=84mm]{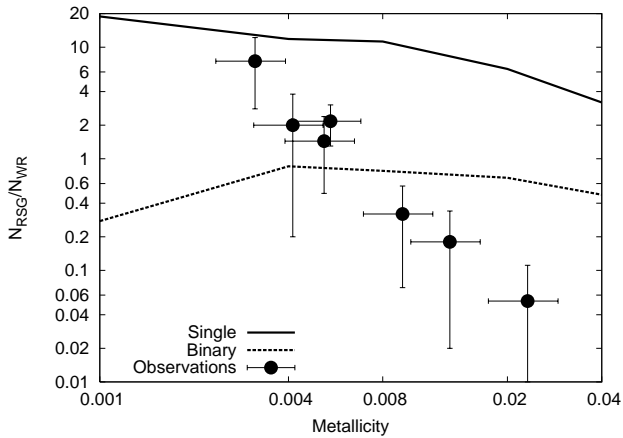}
\caption{Ratio of the number of RSG to WR stars versus metallicity. The observations are taken from \citet{massey}. The solid line is from our single star models while the dashed line is from our binary models.}
\label{rsg2wr}
\end{figure}

In Table \ref{red2wrtable} we list the observed populations. We plot the ratio in Figure \ref{rsg2wr}, it decreases dramatically with increasing metallicity. Such a change would require a stronger scaling of mass loss with metallicity than we currently employ.

Our single star models agree with the observations at SMC metallicity while our binary models agree with the observations around LMC metallicity. One way to match the observed trend between these metallicities is to have a metallicity dependent binary fraction. However something extra is still required around solar metallicity to get an exact agreement.
%%%%VANBEV???
At the higher metallicities the ratios are based on a small number of observed stars (see Table \ref{red2wrtable}) and therefore do not sample the full binary parameter space. Furthermore, small numbers mean our assumptions of constant star formation and IMF become invalid. For example, if all the WR stars are in very close binaries the RSG/WR ratio would be even smaller than we estimate. A large number of stars must be observed to calculate the population ratio to ensure the full range of possible binary systems is probed.

\begin{table}
\caption{Observed numbers and population ratios of RSG and WR stars. The values are taken from \citet{massey}.}
\label{red2wrtable}
\begin{tabular}{@{}lcccc@{}}
\hline
System	&$\log[{\rm O/H}]+12$	&$N_{\rm RSG}$	&$N_{\rm WR}$	&$N_{\rm RSG}/N_{\rm WR}$			\\	
\hline
SMC	    &8.13		&90	&12	&$7.5 \pm 4.7$		\\
NGC6822	&8.25		&8	&4	&$2.0 \pm 1.8$			\\
M33(o)	&8.35		&26	&18	&$1.44 \pm 0.95$	\\
LMC	    &8.37		&234	&108	&$2.17 \pm 0.87$	\\
M33(m)	&8.6		&7	&22	&$0.32 \pm 0.25$	\\
M33(i)	&8.75		&3	&17	&$0.18 \pm 0.16$	\\
M31	    &9		&1	&19	&$0.053 \pm 0.058$	\\
\hline
\end{tabular}
\end{table} 

\subsection{WC to WN Ratio}

\begin{figure}
\includegraphics[angle=270, width=84mm]{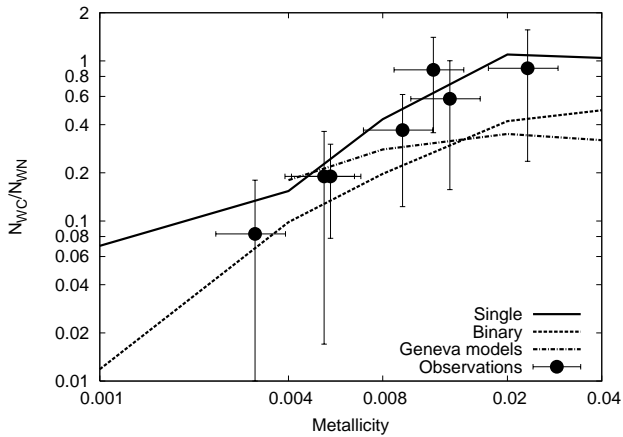}
\caption{Ratio of the number of WC to WN stars versus metallicity. The observations and the dashed-dotted line for the Geneva models are taken from \citet{mm2005}. The solid line is from our single star models while the dashed line is from our binary models.}
\label{wc2wn}
\end{figure}

\begin{table}
\caption{Observed numbers and population ratios of WC and WN stars. The values are taken from \citet{mm2005}.}
\label{wc2wntable}
\begin{tabular}{@{}lcccc@{}}
\hline
System	&$\log[{\rm O/H}]+12$	&$N_{\rm WC}$	&$N_{\rm WN}$	&$N_{\rm WC}/N_{\rm WN}$		\\	
\hline
SMC	    &8.13			&1	&11	&$0.083 \pm 0.10$	\\
NGC6822	&8.25			&0	&4	&$0.000 \pm 0.20$  \\
M33(o)	&8.35			&3	&16	&$0.190 \pm 0.17$	\\
LMC	    &8.37			&17	&91	&$0.190 \pm 0.11$	\\
M33(m)	&8.6			&12	&35	&$0.370 \pm 0.25$\\
MW($<3$kpc)&8.7		&30	&35	&$0.880 \pm 0.52$\\
M33(i)	&8.75		&11	&19	&$0.580 \pm 0.42$		\\
M31	    &9			&13	&14	&$0.900 \pm 0.66$	\\
\hline
\end{tabular}
\end{table} 

Finally, we consider a ratio which provides a measure of the relative lifetimes of the two main WR star types: WN and WC (we include WO with WC). The lifetimes of the two types are determined by the mass-loss rates of WR stars.

We plot the observations and predicted ratios in Figure~\ref{wc2wn} it shows our predicted ratios and the most recent Geneva group rotating models \citep{mm2005} for comparison. The trend of the observations is for the ratio to decrease with metallicity. In agreement with the results of \citet{evink} we find that the single-star models that scale the mass-loss rate of WR stars with initial metallicity agree with the observed trend.

The binary models in this case give a lower WC/WN ratio than the single star models. This is because, as can be seen in Figures \ref{RSGS} and \ref{WNLS}, the lifetimes of WN stars increase by a greater factor than the lifetimes of WC stars in binaries. If we combine a population of single and binary stars the resulting ratio is too low and requires an increase in the WC population relative to the WN population to regain an improved agreement. The result is similar to that found by \citet{vanbevwr}.

\section{Supernova progenitors}

There are many studies which investigate the connection between SNe and massive stars. Some studies consider single-star evolution and predict the initial parameter space and the relative rate of different SN types \citep{H03,ETsne,hmm2004}, while other studies are concerned with the evolution of binary-stars \citep{podsibin1,vanb03,izzysne}. In this section we use our single- and binary-star models to predict the relative SN rates and determine how they vary with metallicity. Firstly, we link our models to each SN type then, secondly, we predict how the relative SN rates vary with metallicity.

Core-collapse SNe are classified according to their lightcurve shapes and spectra. Matching stellar models to observed SNe is difficult and example schemes can be found in \citet{H03} and \citet{ETsne}. Here we check the amount of hydrogen in the progenitor model: if there is more than 0.001M$_{\odot}$ of hydrogen left in the stellar envelope the SN is of type II, otherwise type Ib/c.

There are many subtypes of SNe. The main distinguishing criterion is the mass of the hydrogen or helium envelope at the time of explosion but sometimes the circumstellar environment must also be considered \citep{H03,ETsne}. In this paper we group SN II sub-types (e.g. P, L) together as type II and SN Ib and Ic together as type Ib/c. In our single-star models the initial mass range of Ib/c progenitors is restricted to the most massive stars \citep{ETsne}. In our binary-star models the full range of masses can lead to type Ib/c SNe.

SN rates in different galaxy types have been measured for some time \citep{capp,capp2} and, more recently, SNe observations have been used to determine how the relative rate of type Ib/c to type II SNe varies with metallicity \citep{snevsZ}. The errors especially in the absolute rates of the searches are considerable owing to the small sample size and the uncertainties in the completeness. The relative rates are less uncertain as the selection effects are of similar magnitude and cancel.

\begin{figure}
\includegraphics[angle=270, width=84mm]{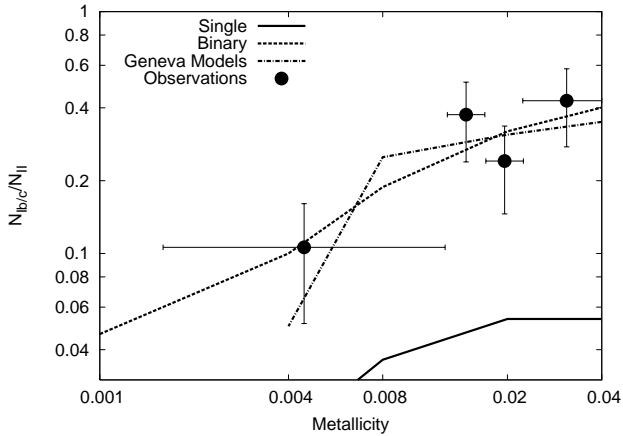}
\caption{The observed and predicted ratios of the type Ib/c SN rate to the type II SN rate. Observations are taken from \citet{snevsZ}. The Geneva model predictions are taken from \citet{mm2005}, the upper line is for their rotating models while the lower line is from their non-rotating models.}
\label{typeIbc2II}
\end{figure}

We plot our predicted SN rate ratios against the observed ratios in Fig. \ref{typeIbc2II}. The observations indicate a general trend of a decreasing rate of type Ib/c SNe relative to type II as metallicity decreases. This is as expected owing to the decreasing strength of stellar winds with reduced metallicity meaning that fewer stars lose all the hydrogen before core-collapse.

We find that our theoretical predictions agree with the trend indicated by the observations. However the single star models predict a lower relative rate for type Ib/c SNe than the observations and the binary star models predict a value that agrees with the observations. We also plot the SN rate ratio predicted by the Geneva rotating models \citep{mm2004} which agree with observations. The conclusion we draw is that we must consider rotating models and/or binary star models to exactly simulate the observations.

\section{Discussion \& Conclusions}
We have compared our stellar models of massive stars to several observations and find that by including binary stars we achieve better agreement. This is because duplicity increases the chance of a star losing its hydrogen envelope which leads to fewer RSGs, more WR stars and more type Ib/c SNe than single star models predict as required by observations.

Problems remain and the agreement is not perfect. The BSG/RSG ratio and RSG/WR ratio indicate that we still predict too many RSGs or not enough WR stars. Also the binary models predict to many WN stars and too few WC stars. Physics we have not included in our models might resolve these problems. The main culprit is rotation. It has two effects on our predictions. First it introduces mixing that extends the time stars spend on the main sequence and therefore increases the number of BSGs. Second, if rotation is rapid it enhances mass loss by reducing the depth of the potential from which mass must escape and therefore increases the population of WR stars at the expense of the RSG population. Rotationally enhanced mass loss affects the WC/WN ratio by shortening the WN lifetimes. However the effect of rotation can be much more complex then these simple effects. For example \citet{mm2001} find that the BSG to RSG ratio is decreased rather than increased by rotation owing to a change of the internal helium gradient at the hydrogen-burning shell. In addition tides in a binary may boost the importance of rotation during important phases of evolution and complicate the situation further \citep{petrovic2,petrovic}. 

Another limitation of our models is that we do not include mass losing eruptions that are common in the luminous-blue variable (LBV) stars. These most massive stars experience giant outbursts, losing large amounts of mass in one short event. This has dramatic implications for the evolution of the object \citep{solbv}. There is growing evidence that SN progenitors can experience these outbursts prior to the SN explosion \citep{kvlbv,06jc}. Such outbursts would have a similar effect on stellar populations as rotation. They reduce the time stars spend as RSGs and could reduce the time taken to remove the helium envelopes for WN stars. Therefore a better understanding of these outbursts may also improve agreement between stellar models and observations.

In summary, binary stars must be considered when comparing stellar evolution models to observations. However, binary evolution alone cannot explain all the observations. Fine tuning of stellar models is still required and the effect of enhanced mass loss owing to rotation or LBV-like eruptions must be further considered.

%Finally find that a binary companion can have an important effect on the mass lost by a star and affects the eventual SN type. This is only the case if the star is below the initial single-star mass-limit for a type Ib/c SN ($\la$ 27M$_{\odot}$ at solar metallicity). If the star is massive enough that stellar winds can remove the hydrogen envelope, these stars will always result in a type Ib/c SN. The stars below the initial mass-limit for a type Ib/c SN that would normally retain their hydrogen envelopes as single stars are therefore the ones that will give rise to the other SN types, with the final type of SN determined by the strength of the binary interaction during the evolution.

\section{Acknowledgements}
The authors would like to thank the referee Andre Maeder for his helpful comments that improved the paper. JJE conducted part of this work during his time at the IAP in France as a CRNS post-doc. The remainder of the work was carried out as part of the award ``Understanding the lives of massive stars from birth to supernovae'' made under the European Heads of Research Councils and European Science Foundation EURYI Awards scheme and was supported by funds from the Participating Organisations of EURYI and the EC Sixth Framework Programme. CAT thanks Churchill College, Cambridge for his Fellowship. RGI thanks the NWO for his current fellowship in Utrecht. JJE also thanks Stephen Smartt, Norbert Langer and Onno Pols for useful discussion.

\bsp

\end{document}